\documentclass[pra,aps,a4paper,twocolumn]{revtex4}

\usepackage{bm}
\usepackage{amsmath}
\usepackage{amsfonts}
\usepackage{amssymb}
\usepackage{color}
\usepackage{graphicx}
\usepackage{changes}
\usepackage[final]{pdfpages}
\usepackage{pgffor}

\makeatletter
\AtBeginDocument{\let\LS@rot\@undefined}
\makeatother

\begin{document}

\title{Current noise geometrically generated by a driven magnet}


\author{Tim \surname{Ludwig}$^{1,2}$, Igor S. \surname{Burmistrov}$^{2,3,1,4}$, Yuval \surname{Gefen}$^5$, 
Alexander \surname{Shnirman}$^{1,4}$}
\affiliation{$^1$Institut f\"ur Theorie der Kondensierten Materie, Karlsruhe Institute of Technology, 76128 Karlsruhe, Germany}
\affiliation{$^2$L.D. Landau Institute for Theoretical Physics RAS, Kosygina street 2, 119334 Moscow, Russia}
\affiliation{$^3$Laboratory for Condensed Matter Physics, National Research University Higher School of Economics, 101000  Moscow, Russia}
\affiliation{$^4$Institute of Nanotechnology, Karlsruhe Institute of Technology, 76021 Karlsruhe, Germany}
\affiliation{$^5$Department of Condensed Matter Physics, Weizmann Institute of Science, 76100 Rehovot, Israel}

\begin{abstract}
We consider a non-equilibrium cross-response phenomenon, whereby a driven magnetization gives rise to electric shot noise (but no d.c. current). This effect is realized on a nano-scale, with a small metallic ferromagnet
which is tunnel-coupled to two normal metal leads. The driving gives rise to a precessing magnetization.  The geometrically generated noise is related to a non-equilibrium distribution in the ferromagnet. Our protocol  provides a new channel for detecting and characterizing ferromagnetic resonance.
\end{abstract}

\maketitle

Off-diagonal (cross-) response phenomena, e.g. the thermoelectric effect, are ubiquitous in physics. In spintronic systems, by applying an electric charge current one can drive magnetization dynamics and vice versa~\cite{slonczewski1996current, PhysRevB.54.9353, PhysRevB.59.11465, PhysRevLett.88.117601, PhysRevB.66.060404, RevModPhys.77.1375, PhysRevB.78.020401}. This usually requires magnetic contacts which allow for a conversion between spin and charge currents; see however \cite{jonson2019dc}.
In this Letter we report a higher order strongly non-equilibrium cross-response effect. Namely, we show that by driving magnetization dynamics one can generate electric shot noise \cite{landauer1998condensed, blanter2000shot} without generating charge current.  Strikingly, no magnetic leads are needed and the leads can be at equilibrium with each other.

\begin{figure}[b]
\begin{center}
\includegraphics[width=0.4\textwidth]{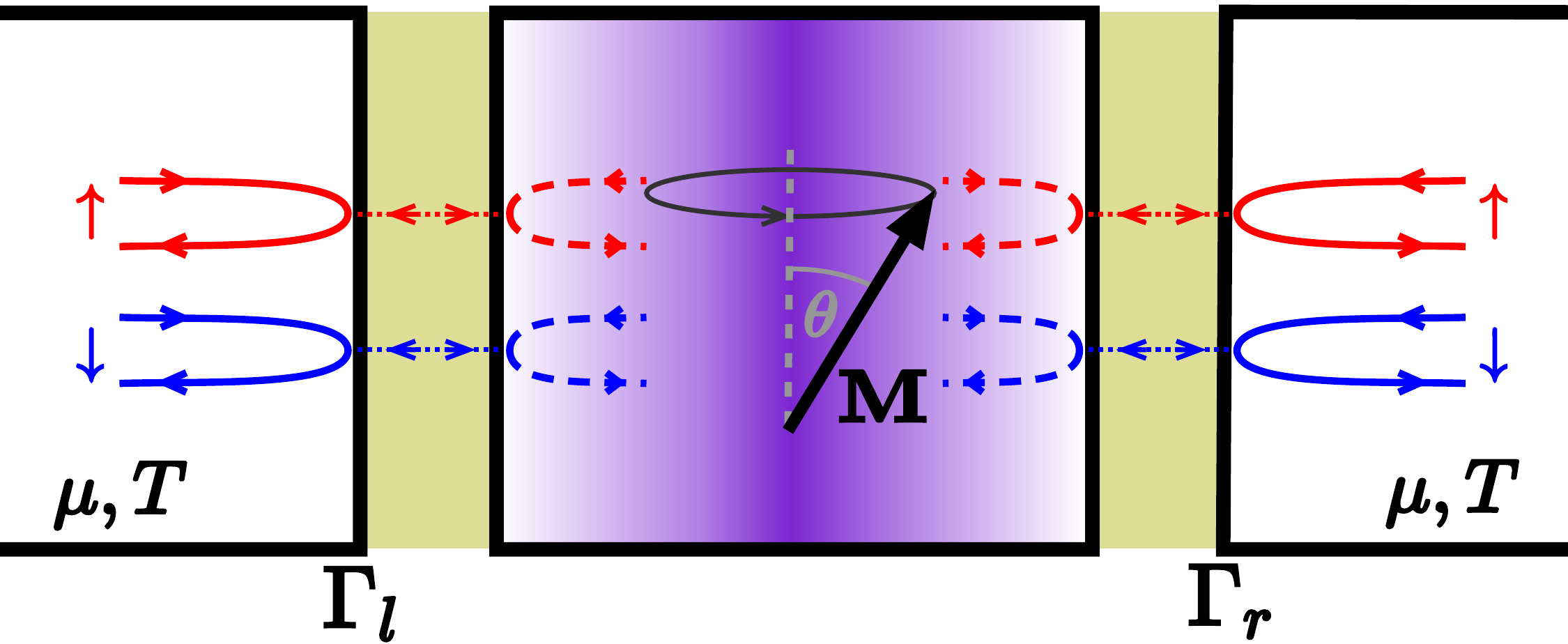}
\end{center}
\caption{A small metallic ferromagnet with precessing magnetization is tunnel-coupled to two normal metal leads, which are at equilibrium with each other. 
The precessing magnetization pumps a spin-current from the small ferromagnet into the leads \cite{PhysRevLett.88.117601, PhysRevB.66.060404, RevModPhys.77.1375}. The average charge current vanishes by symmetry.
Thus, the current of spin-up electrons and spin-down electrons balance each other on average and in each junction separately; in the ferromagnet, the precessing magnetization mixes spin-up and spin-down electrons. All four spin-resolved electron currents are fluctuating.
These fluctuations combine to give rise to the noise of left to right (transport) charge current.} \label{fig: system}
\end{figure}

We consider a small metallic ferromagnet with magnetization driven to precess. The ferromagnet is tunnel-coupled to two normal metal leads; see Fig. \ref{fig: system}.  
The precessing magnetization drives the electrons of the ferromagnet into a strongly non-equilibrium state. This effect is most pronounced if the ferromagnet is small enough such that internal relaxation is negligible compared to the relaxation due to the coupling to the leads. The precessing magnetization, in turn, induces non-equilibrium shot noise of the electric current. The non-equilibrium distribution responsible for the shot noise is governed by the geometric Berry phase due to precessing magnetization, branding the shot noise geometric. This shot noise exists even when both leads are in equilibrium with each other, although the average charge current vanishes then.

Shot noise is particularly interesting in spintronics because it gives insights into the magnetic configuration and its dynamics which may be hard to obtain otherwise \cite{PhysRevB.79.214407, PhysRevLett.114.016601, PhysRevB.94.014419, PhysRevLett.116.146601, cascales2015detection, PhysRevLett.118.237701, doubledot}.

\begin{figure}
\begin{center}
\includegraphics[width=0.425\textwidth]{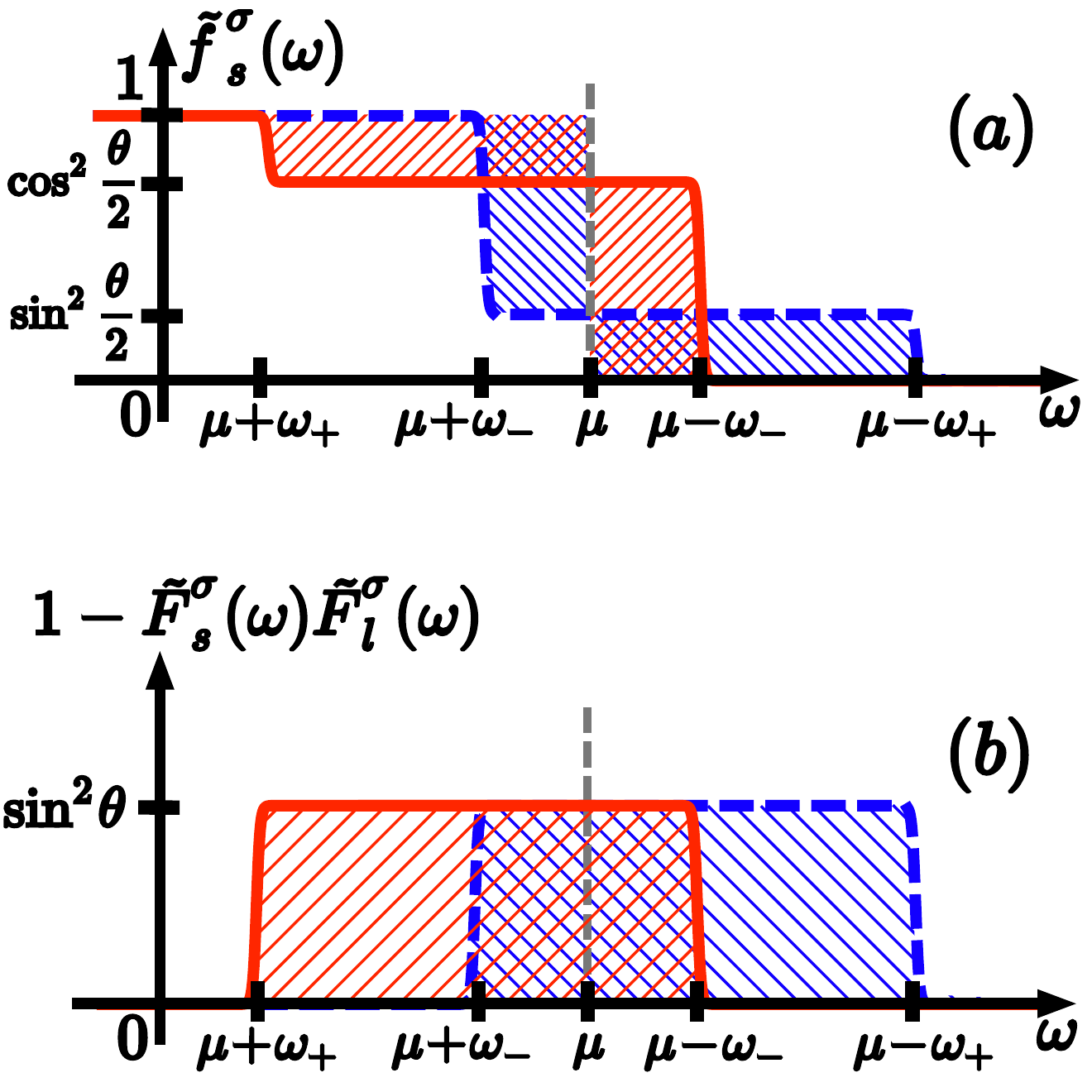}
\end{center}
\caption{(a) The spin-diagonal part of the small ferromagnet's \textit{rotating-frame} distribution function is shown for spin-up (red solid) and spin-down (blue dashed). 
The areas shaded in blue and red are all equal in size: $\sin^2 \theta\ |\dot \phi|/4$, which means that the electrons are redistributed in energy space for each spin-polarization separately. (b) The noise of charge current, Eq. \eqref{eq: noise}, is determined by the integral over $1 - \tilde F_s^\sigma(\omega) \tilde F_l^\sigma(\omega)$, which itself is governed by the distribution function $\tilde f_s^\sigma(\omega)$. The contribution to this noise is identical for both spin-polarizations, as the shaded areas are equal in size (red for spin-up; blue for spin-down). Parameters in figures: $\theta = \pi/3$ and  $\omega_d < 0$.}\label{fig: distributions}
\end{figure}

\textit{Results.}---In order to describe dynamics of the magnetization of a small ferromagnet we use the macrospin approximation, i.e.,  the magnetization is given by a single vector $\bm M = M (\sin \theta \cos \phi, \sin \theta \sin \phi, \cos \theta)$. We assume a steady state precession of the magnetization at a constant polar angle $\theta$ and with a constant precession frequency $\dot\varphi$. Under this assumptions, we found that the charge current vanishes on average, $I  = 0$, but the current noise remains finite:
\begin{equation}
S =   4 g_t T + g_t \sin^2 \theta\, \Big( \dot \phi\, \coth{\frac{\dot \phi}{2T}} - 2T\Big)\ .\label{eq: shot noise}
\end{equation}
Here $g_t = 2 (\rho_\uparrow + \rho_\downarrow) \Gamma_l \Gamma_r/ (\Gamma_l + \Gamma_r)$ is the total conductance of the double tunnel-junction with spin-dependent density of states of the small ferromagnet $\rho_{\sigma}$. The rates $\Gamma_l$ and $\Gamma_r$ characterize the spin-conserved tunneling to left and right leads respectively. 
The precessing magnetization pumps a spin-current into the adjacent leads \cite{PhysRevLett.88.117601, PhysRevB.66.060404, RevModPhys.77.1375}, which drives the electron system into a strong non-equilibrium state \cite{PhysRevB.95.075425, PhysRevB.99.045429}; see Fig. \ref{fig: distributions}.
At high temperatures $(T \gg \dot \phi)$, the noise is dominated by the first term $S \approx 4 g_t T$, which is the standard thermal noise.
At low temperatures $(T \ll  \dot \phi\, \sin^2 \theta)$, however, the noise is dominated by the second term $S \approx g_t \sin^2 \theta\, |\dot \phi|$.
The time-dependence of the magnetization is the source of driving for the electron system. Therefore, the precession frequency $\dot \phi$ acts like a voltage bias for standard shot noise.

\begin{figure}
\begin{center}
\includegraphics[width=0.45\textwidth]{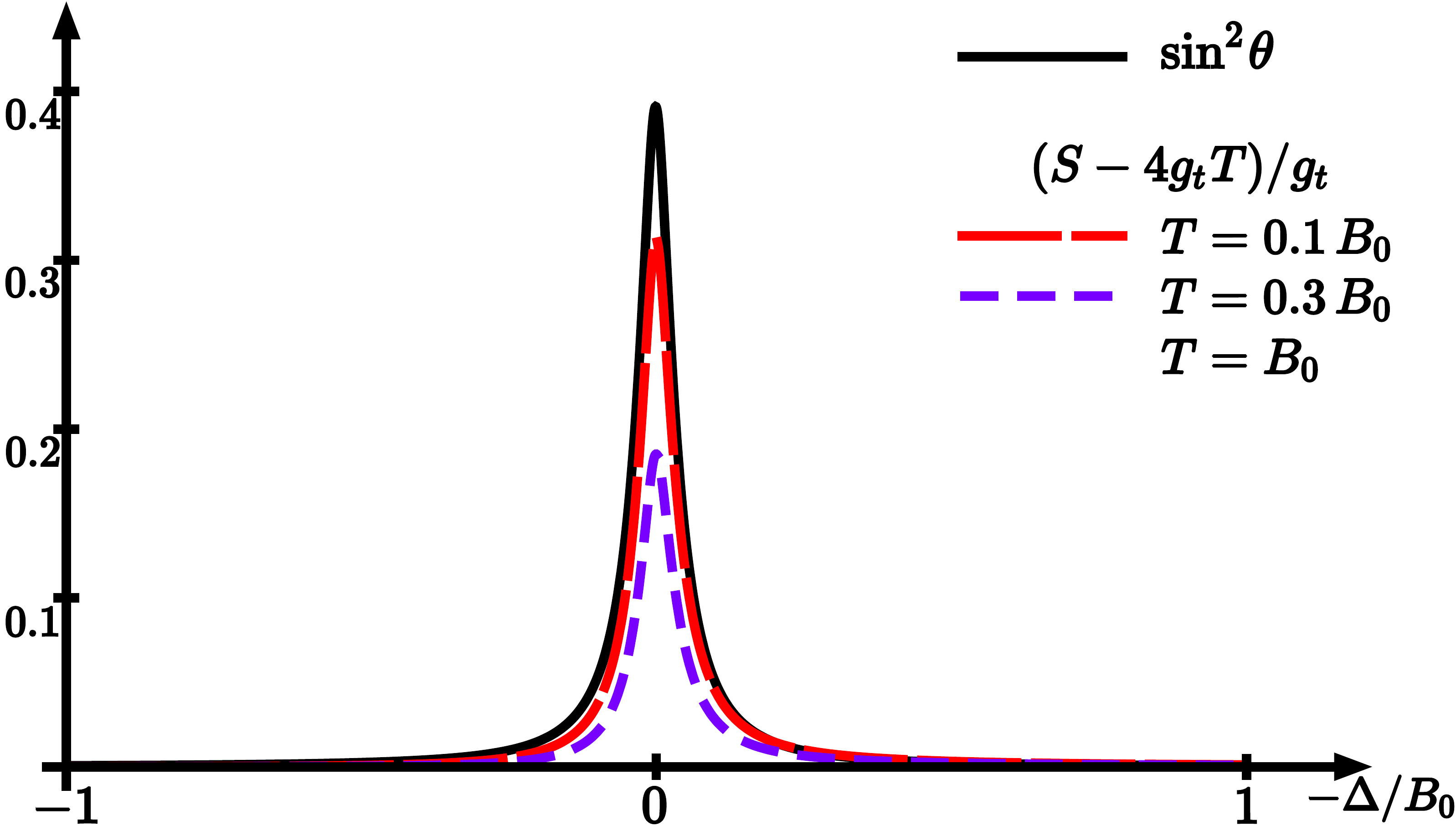}
\end{center}
\caption{When the steady state precession of the magnetization is maintained by driving with a FMR-setup, the polar angle $\theta$ depends on driving frequency $\dot \phi = \omega_d$. The peak of $\sin^2 \theta$ at $\omega_d = - B_0$ ($\Delta = 0$) is a typical FMR-peak. We show the zero-frequency noise of charge current that is generated by the precessing magnetization; we subtract the thermal contribution and normalize onto the value of the total conductance, that is, we show $(S - 4 g_t T)/g_t$. The generated noise of charge current clearly reflects the peak structure of $\sin^2 \theta$ in the FMR-setup. Parameters in figure: $\alpha = 0.04$, $\Omega/(\alpha B_0) = 0.63$.} \label{fig: noise}
\end{figure}

\textit{Application to FMR-driven magnet.}---Now let us consider our setup under conditions of a ferromagnetic resonance (FMR).
The dynamics of the magnetization is phenomenologically described by the Landau-Lifshitz-Gilbert equation $\bm{\dot{m}} = \bm m \times \bm B - \alpha\, \bm m \times \bm{\dot{m}}$, where $\bm m = \bm M / M$ is the direction of the magnetization and $\alpha$ is the Gilbert-damping coefficient. For the FMR-setup, we choose the magnetic field $\bm B = (\Omega \cos \omega_d t, \Omega \sin \omega_d t, B_0)$ with a fixed component $B_0$ in $z-$direction and, perpendicular to it, a small driving field with strength $\Omega$ and frequency $\omega_d$. 
For negligible internal relaxation, the damping is dominated by the coupling to the leads.
Without driving, the Gilbert-damping would relax the magnetization towards $\theta = 0$.
With driving ($\Omega \neq 0$), however, the magnetization can be brought into a steady state precession. That is, after the decay of transient effects, the magnetization precesses at the frequency of the driving field $\dot \phi = \omega_d$ and the polar angle $\theta$ is determined by the competition between Gilbert-damping and FMR-driving. Explicitly, $\theta$ is determined by 
\begin{equation}
\sin^2\theta = \frac{(\Omega_++\Omega_-)^2}{\Omega^2_++\Omega^2_-+2\Delta^2+2\sqrt{(\Delta^2+\Omega_+^2)(\Delta^2+\Omega_-^2)}} \ ,
\end{equation}
with $\Omega_\pm = \Omega \pm \alpha \omega_d$ and the detuning parameter $\Delta=\omega_d+B_0$.
The dependence of $\sin^2\theta$ on precession frequency $\omega_d$ has a resonant character with a maximum at $\omega_d=-B_0$. This ferromagnetic resonance of the magnetization's steady state precession directly translates into a resonance in the current noise; see Fig. \ref{fig: noise}. At low temperatures, $T\ll \omega_d \sin^2 \theta$, the form of the resonance in the current noise resembles the FMR structure of the stationary precession angle. At higher temperatures, the resonance in the current noise can be visible on top of the constant thermal noise.
Now we explain how our results were derived.

\textit{The effective action.}---Because the dynamics of the magnetization creates non-equilibrium conditions, we apply Keldysh formalism \cite{KamenevBook, doi:10.1080/00018730902850504, altland2010condensed}. The Keldysh generating function is $\mathcal Z = \int D[\bar \Psi, \Psi]\, \exp{(i \mathcal{S})}$ with the action,
\begin{equation}
\mathcal{S} = \oint_K\!\! dt\ \bar \Psi\, (i \partial_t - h_s - \hat \Sigma )\, \Psi\ , \label{eq: action}
\end{equation}
where the integral is along the Keldysh contour and $\Psi, \bar \Psi$ denote the fermionic fields of the small ferromagnet. The self-energy operator $\hat \Sigma$ is defined by $[\hat \Sigma \Psi](t) = \oint dt'\, \Sigma(t-t') \Psi(t')$, where $\Sigma = \Sigma_l + \Sigma_r$ is the self-energy arising from the tunnel-coupling to the left lead $\Sigma_l$ and right lead $\Sigma_r$. 
The self-energy contains the essential information about the tunnel-coupling to the leads: 
first, the retarded and advanced part contain the tunneling-rates $\Sigma^{R/A}_{l/r} (\omega) = \mp i \Gamma_{l/r}$; 
second, the Keldysh part contains the distribution functions of the leads $\Sigma^K_{l/r}(\omega) = - 2i \Gamma_{l/r} F_{l/r}(\omega)$, where $F_{l/r}(\omega) = 1- 2 f_{l/r}(\omega)$ with the Fermi-distributions $f_{l/r}(\omega) = 1/[\exp [ (\omega-\mu )/T] + 1]$. 
We emphasize that the ferromagnet's distribution function $f_s$, respectively $F_s$, is not yet known explicitly but it is implicitly determined by the action, Eq. \eqref{eq: action}. This distribution function is governed by the coupling to the leads \textit{and} the dynamics of the magnetization which enters through the effective single-particle Hamiltonian,\begin{equation}
h_s = h_0 -  \bm M \bm{\sigma}/2\, \label{eq: single particle Hamiltonian}
\end{equation}
where $\bm \sigma$ is the vector of Pauli-matrices and $h_0$ is a spin-degenerate single-particle Hamiltonian of the small ferromagnet.
For the derivation of the charge noise, the magnetization is considered to be a classical field with given dynamics (steady state precession).

The charge current and its noise are determined with help of a counting field $\lambda$, which is introduced into the self-energy related to the left lead $\Sigma_l \rightarrow \Sigma_l(\lambda)$. We follow Ref. \cite{PhysRevLett.118.237701}, and introduce $\lambda$ such that the charge transported through the left junction is determined as $\left\langle Q_l \right\rangle = i \partial_\lambda \left. \mathcal Z(\lambda)\right|_{\lambda=0}$ with the corresponding noise $\left\langle Q_l^2 \right\rangle = (i \partial_\lambda)^2 \left. \mathcal Z(\lambda)\right|_{\lambda=0}$; details are provided in supplementary material (SM).
We can now integrate out the fermions to obtain $\mathcal{Z}(\lambda) = \exp [i \mathcal{S}(\lambda)]$ with the action
\begin{equation}
i\mathcal{S}(\lambda) = \mathrm{tr}\, \mathrm{ln} \big[ i \partial_t - h_0 + \bm{M} \bm{\sigma}/2 - \Sigma(\lambda) \big]\ . \label{eq: action fermions integrated out}
\end{equation}
The magnetization's time-dependence makes it complicated to proceed. It is, thus, very convenient to transform to a frame of reference in which the magnetization is time-independent.

\textit{Rotating frame.}---The magnetization is rotated onto the $z$-axis at all times, 
\begin{equation}
R^\dagger \bm M \bm \sigma R = M \sigma_z\ ,
\end{equation}
with a time-dependent rotation in spin-space $R$.
While simplifying the magnetic part, this rotation also comes at a cost: because of its time-dependence, it gives rise to a new term $i R^\dagger \dot R$ under the $\mathrm{tr}\, \mathrm{ln}$, see Eq. \eqref{eq: action fermions integrated out}, and also rotates the self-energy $R^\dagger \Sigma R$. After rotation, the action becomes
\begin{equation}
i \mathcal{S}(\lambda) = \mathrm{tr}\, \mathrm{ln} \big[ \underbrace{i \partial_t - h_0 + M \sigma_z/2 + i R^\dagger \dot R - R^\dagger \Sigma(\lambda) R}_{=: \tilde G^{-1}_\lambda}\big]\ , \label{eq: rotated action fermions integrated out}
\end{equation}
where $\tilde G^{-1}_\lambda$ defined the rotating-frame Green's function $\tilde G_\lambda$.
Following Ref. \cite{PhysRevLett.114.176806}, we choose the Euler-angle representation $R= e^{-i \frac{\phi}{2} \sigma_z} e^{-i \frac{\theta}{2} \sigma_y} e^{i \frac{\phi - \chi}{2} \sigma_z}$, 
where $\phi, \theta$ are the angles characterizing the magnetization and the gauge-freedom $\chi$ is fixed by $\dot \chi = \dot \phi (1- \cos \theta)$. This choice eliminates the spin-diagonal part of $iR^\dagger \dot R$ which contains information about the Berry phase. However, the Berry phase is not eliminated; instead it is shifted to the rotated self-energy.

\textit{Rotating-frame distribution functions.}---Because retarded and advanced parts of the self-energy are trivial in spin-space and local in time, the rotation only affects the Keldysh part. While the Keldysh part $\Sigma^K(t-t') = - 2i [ \Gamma_l  F_l(t - t') + \Gamma_r  F_r(t - t')]$ is also trivial in spin-space, it is non-local in time because of the distribution functions $F_{l/r}(t-t')$. It follows, $R^\dagger(t) \Sigma^K(t-t') R(t') = - 2i [ \Gamma_l \tilde F_l(t,t') + \Gamma_r \tilde F_r(t,t')]$ with the \textit{rotating-frame distribution functions} $\tilde F_{l/r}(t,t') = R^\dagger(t) F_{l/r}(t-t') R(t')$. 
For the following, it is convenient to change to the Wigner time-coordinates $\bar t = (t + t')/2$, $\Delta t =  t -t'$ and to perform a Fourier-transformation $\Delta t \rightarrow \omega$.
For steady state precessions (with $\theta$ and $\dot \phi$ constant), the spin-diagonal parts of the rotating-frame distribution functions are given by $\tilde F_{l/r}^\sigma(\omega )= [ \tilde F_{l/r}(\omega)]_{\sigma \sigma}= \cos^2 \frac{\theta}{2} F_{l/r}(\omega + \sigma \omega_-) + \sin^2 \frac{\theta}{2} F_{l/r}(\omega + \bar \sigma \omega_+)$. These distributions are governed by the magnetization dynamics via the Berry-phase in $\omega_\pm = \dot \phi (1 \pm \cos \theta)/2$.

\textit{Adiabatic approximation.}---In order to proceed, we have to determine the rotating-frame Green's function $\tilde G_\lambda$ for vanishing counting field $\lambda = 0$.
In principle this poses a complicated problem, since the spin-off-diagonal elements of its inverse $\tilde G_0^{-1}$ depend on time.
However, we assume the magnetization $M$ to be the largest relevant energy scale in the small ferromagnet. This allows us to disregard the spin-off-diagonal elements of $\tilde  G_0^{-1}$  for the determination of $\tilde G_0$
In particular, we disregard spin-off-diagonal elements of $i R^\dagger \dot R$ which are related to transitions between spin-up and spin-down states; this corresponds to an adiabatic approximation \cite{PhysRevLett.114.176806}. Furthermore, we disregard spin-off-diagonal elements of the rotated self-energy. It is, now, straightforward to obtain the rotating-frame Green's function,
\begin{equation}
\begin{split}
\tilde G_{0,a \sigma}^{R/A}(\omega) &= \frac{1}{\omega - \xi_{a \sigma} \pm i \Gamma_\Sigma}\ , \\
\tilde G_{0,a \sigma}^K(\omega) &= \frac{-2i \Gamma_\Sigma \tilde F_s^\sigma(\omega)}{(\omega - \xi_{a \sigma})^2 + \Gamma_\Sigma^2}\ ,
\end{split}
\end{equation}
with the total level broadening $\Gamma_\Sigma = \Gamma_l + \Gamma_r$. The spin-dependent single-particle energy is $\xi_{a \sigma} = \epsilon_a - M \sigma/2$, where $\epsilon_a$ are the eigenenergies of $h_0$ with corresponding eigenstates $a$. The rotating-frame distribution function of the small ferromagnet, 
\begin{equation}
\tilde F_s^\sigma(\omega) = \big[ \Gamma_l \tilde F_l^\sigma(\omega) + \Gamma_r \tilde F_r^\sigma(\omega) \big] / \Gamma_\Sigma \ , \label{eq: small fm distribution}
\end{equation}
is a superposition of the leads' rotating-frame distribution functions. In absence of bias, the rotating-frame distribution functions are exactly the same in all three systems $\tilde F_s^\sigma(\omega)= \tilde F_l^\sigma(\omega) = \tilde F_r^\sigma(\omega)$; see Fig. \ref{fig: distributions}.

It is worthwhile to emphasize that the transformation into the rotating frame is a crucial step that allows us to solve the problem. The reason is as follows. As we discussed above it is enough to find the spin-diagonal components of the rotating-frame distribution function. However, as one can check \cite{SM}, the knowledge of the spin-diagonal components of the rotating-frame distribution function is not enough in order to determine the distribution function in the laboratory frame.

\textit{Charge current and its noise.}---The zero-frequency charge current $I_l$ is defined via the transported charge $\langle Q_l \rangle = \int dt\, I_l$. Differentiating the generating function, the transported charge is determined to $\langle Q_l \rangle = - i\, \mathrm{tr} [ \tilde G_0 \tilde \Sigma_l']$, where $\tilde \Sigma_l' = \partial_\lambda \tilde \Sigma_l(\lambda)|_{\lambda = 0}$ is the derivative of the rotated self-energy $\tilde \Sigma(\lambda) = R^\dagger \Sigma (\lambda) R$. For the current, we find \cite{SM}
\begin{equation}
I_l =\ \sum_\sigma \rho_\sigma \Gamma_l \int d \omega\,  [\tilde F^\sigma_l(\omega)  - \tilde F_s^\sigma(\omega) ]\ = 0\ , \label{eq: current}
\end{equation}
where we defined the spin-dependent density of states, $\rho_\sigma(\omega) = \sum_a \frac{1}{\pi} \frac{\Gamma_\Sigma}{(\omega - \xi_{a \sigma})^2 + \Gamma_\Sigma^2}$. We assumed it to be approximately constant $\rho_\sigma(\omega) = \rho_\sigma$ on all scales smaller than $M$. The resulting formula for the charge current is the Landauer formula \cite{landauer1957spatial} with rotating-frame distribution functions. This reflects the fact that the amount of transported charge is an observable which has to be independent of the frame of reference. Explicitly, however, the current vanishes, since no bias is applied.

Similar to the average current, the zero-frequency noise \footnote{Strictly speaking, it should be called low-frequency noise, as fluctuations of the magnetization become important for measurement times $t_m$ longer that the typical relaxation time $\tau_\theta$ of the $\theta$ coordinate \cite{PhysRevLett.118.237701}. This relaxation time, however, scales with the magnetization length $\tau_\theta \sim M$ which we assumed to be very large.} of charge current $S_l$ is defined via $\langle \langle Q_l^2 \rangle \rangle = \int dt\, S_l/2$. Differentiating the generating function, the noise of transported charge is determined to $\langle \langle Q_l^2 \rangle \rangle = \mathrm{tr} [ \tilde G_0 \tilde \Sigma_l''] + \mathrm{tr} [ \tilde G_0' \tilde \Sigma_l']$, where $\tilde \Sigma_l'' = \partial^2_\lambda \tilde \Sigma_l(\lambda)|_{\lambda = 0}$ is the second derivative of the rotated self-energy and $\tilde G_0' = \partial_\lambda  \tilde G_\lambda \big|_{\lambda = 0} = \tilde G_0 \tilde \Sigma_l' \tilde G_0$ is the derivative of the rotating-frame Green's function. For the noise, we find \cite{SM}
\begin{align}
S_l &=  \sum_\sigma g_\sigma \int d \omega\, \Big\lbrace \big[1-\tilde F_s^\sigma(\omega) \tilde F_l^\sigma(\omega)\big] + \nonumber \\
& \hspace{7em} + \frac{\Gamma_l}{\Gamma_r} \tilde F_s^\sigma(\omega) \big[ \tilde F_l^\sigma(\omega) - \tilde F_s^\sigma(\omega)\big] \Big\rbrace\ , \label{eq: noise} 
\end{align}
where $g_\sigma = 2 \rho_\sigma \Gamma_l \Gamma_r / (\Gamma_l + \Gamma_r)$ is the spin-dependent conductance of the double tunnel-junction.
After the integration over frequency, we obtain Eq. \eqref{eq: shot noise} as result for the shot noise.

\textit{Discussion.}---
In our relatively simple model which excludes internal relaxation, we were able to properly derive the non-equilibrium 
distribution function $\tilde F_s^\sigma(\omega)$ given by Eq. \eqref{eq: small fm distribution}. 
This, in particular, guarantees that the charge conservation laws are satisfied. 
Indeed, since the small ferromagnet cannot store additional charges for an infinite time, charge conservation requires $I_l = -I_r =:I$ and $S_l = S_r =:S$ at zero frequency. For the right junction, current $I_r$ and noise $S_r$ can be obtained from eqs. \eqref{eq: current} and \eqref{eq: noise} by exchanging $\Gamma_l \leftrightarrow \Gamma_r$ and substituting $\tilde F_l^\sigma (\omega) \rightarrow \tilde F_r^\sigma (\omega)$. As expected, we find $I_l = -I_r$ and $S_l = S_r$. 

In the presence of internal relaxation one might be tempted to impose a physically motivated distribution function in the small ferromagnet as a shortcut of a full calculation. We emphasize, however, that the charge conservation condition $S_l = S_r$ puts a strong restriction onto possible distribution functions. 
In particular, charge conservation would be violated if $\tilde F_s^\sigma(\omega)$ is just replaced by
an equilibrium distribution function with an adjusted electrochemical potential. Thus, a straightforward application of  the results of Ref. \cite{PhysRevLett.118.237701} obtained for a single tunnel junction to the double tunnel-junction considered here is not possible.

We  expect the effects of internal relaxation to be threefold: (i) the Gilbert-damping coefficient $\alpha$ can be increased (spin-orbit coupling) and, thereby, the polar angle $\theta$ of steady state precessions is changed; (ii) internal relaxation tends to bring the magnet's rotating-frame distribution function $\tilde F_s^\sigma(\omega)$ towards equilibrium; (iii) the formal result for the noise, Eq. \eqref{eq: noise}, has to be changed in order not to violate charge conservation when the distribution function changes. However, for weak internal relaxation, these effects might be taken into account perturbatively and, therefore, we expect our results to be robust against finite but small internal relaxation.

\textit{Conclusion.}---
We have found a higher order non-equilibrium off-diagonal response effect. Namely, we
have shown that zero-frequency shot noise of charge current is generated by a precessing magnetization of a small ferromagnet which is tunnel-coupled to two normal metal leads. This noise, Eq. \eqref{eq: noise}, crucially depends on the electronic distribution function which is in turn geometrically governed by the magnetization dynamics; see Fig. \ref{fig: distributions}.
Thus, the noise of the charge current, Eq. \eqref{eq: shot noise}, is generated
by the precession of the magnetization. For the FMR-setup, Fig. \ref{fig: noise}, this effect can be used to detect 
the magnetization dynamics in spite of the vanishing average current.

\textit{Acknowledgements.}---We thank G. E. W. Bauer, M. Ke\ss ler, and W. Wulfhekel for fruitful discussions. This work was supported by DFG Research Grant No. SH 81/3-1. The research of T.L. is partially supported by the Russian Foundation for Basic Research under the Grant No. 19-32-50005. Furthermore, T.L. acknowledges KHYS of KIT and the Feinberg Graduate school of WIS for supporting a stay at WIS; I.S.B. acknowledges RAS Program Topical problems in low temperature physics, the Alexander von Humboldt Foundation, and the Basic research program of HSE; Y.G. acknowledges the DFG Research Grant RO 2247/11-1 and the Italia-Israel QUANTRA.

\bibliography{reference.bib}



\section*{Supplementary material}

# ONLINE SUPPLEMENTAL MATERIAL
# Current noise geometrically generated by a driven magnet

Tim Ludwig[1,2], Igor S. Burmistrov[2,3,1,4], Yuval Gefen[5], Alexander Shnirman[1,4]
[1]*Institut für Theorie der Kondensierten Materie,
Karlsruhe Institute of Technology, 76128 Karlsruhe, Germany*
[2]*L.D. Landau Institute for Theoretical Physics RAS, Kosygina street 2, 119334 Moscow, Russia*
[3]*Laboratory for Condensed Matter Physics, National Research University Higher School of Economics, 101000 Moscow, Russia*
[4]*Institute of Nanotechnology, Karlsruhe Institute of Technology, 76021 Karlsruhe, Germany and*
[5]*Department of Condensed Matter Physics, Weizmann Institute of Science, 76100 Rehovot, Israel*

In this supplemental material, we present (i) details of the introduction of counting fields, (ii) details of the derivation of results for the charge current and its noise, and (iii) a detailed explanation why the knowledge of spin-diagonal components of the rotating-frame distribution function is not sufficient to determine the laboratory-frame distribution function.

## S.I. DETAILS OF THE DERIVATION FOR CHARGE CURRENT AND ITS NOISE

### A. Derivation of effective action

Before leads are integrated out, the full Hamiltonian is $H = H_s + H_l + H_r$ where $H_s$ describes the small ferromagnet and $H_l, H_r$ describe the left and right lead respectively and include the tunnel-coupling to the small ferromagnet. Explicitly, $H_s = \sum_{a\sigma\sigma'}[h_s]_{a\sigma\sigma'} c_{a\sigma}^\dagger c_{a\sigma'}$, where $c_{a\sigma}^\dagger$ creates ($c_{a\sigma}$ annihilates) a particle in the orbital state $a$ with spin $\sigma$. The single particle Hamiltonian is $h_s = h_0 - \boldsymbol{M}(t)\frac{\boldsymbol{\sigma}}{2}$, where $\boldsymbol{M}(t)$ is the time dependent magnetization of the small ferromagnet, $\boldsymbol{\sigma}$ is the vector of Pauli-matrices, and $h_0 = \sum_{a\sigma} \epsilon_a c_{a\sigma}^\dagger c_{a\sigma}$ is the spin-degenerate single-particle Hamiltonian. The leads are assumed to be non-interacting and the tunnel-coupling is assumed to be spin-conserving. The (coupling to the) left lead is described by $H_l = \sum_{\gamma\sigma} \epsilon_\gamma c_{\gamma\sigma}^\dagger c_{\gamma\sigma} + \sum_{a\gamma\sigma}(t_{l,an} c_{a\sigma}^\dagger c_{\gamma\sigma} + h.c.)$, where $\gamma = (n,k)$ is a collective index for momentum $k$ and transport channel $n$. Respectively, $c_{\gamma\sigma}^\dagger, c_{\gamma\sigma}$ are creation and annihilation operators for electrons in state $\gamma$ with spin $\sigma$ and $\epsilon_\gamma$ is the corresponding energy. The tunneling through the left junction is described by the tunneling-matrix $t_l$. Analogously, $H_r = \sum_{\tilde\gamma\sigma} \epsilon_{\tilde\gamma} c_{\tilde\gamma\sigma}^\dagger c_{\tilde\gamma\sigma} + \sum_{a\tilde\gamma\sigma}(t_{r,a\tilde n} c_{a\sigma}^\dagger c_{\tilde\gamma\sigma} + h.c.)$ describes (the coupling to) the right lead. Now, applying Keldysh formalism leads to the Keldysh partition function $\mathcal{Z} = \int D[\bar\Psi, \Psi] e^{i\mathcal{S}}$ with the action $\mathcal{S} = \oint_K dt\,[\bar\Psi\, i\partial_t\,\Psi - H(\bar\Psi,\Psi)]$, where $\bar\Psi,\Psi$ denote all fermionic fields including those of the leads. Would we directly integrate out the fermionic fields of the leads, we would obtain the action $\mathcal{S} = \oint_K dt\,\bar\Psi(i\partial_t - h_s - \hat\Sigma)\Psi$, where $\bar\Psi, \Psi$ denote only fermionic fields of the small ferromagnet and the self-energy operator $\hat\Sigma$ is defined by $[\hat\Sigma\Psi](t) = \oint dt'\Sigma(t-t')\Psi(t')$; compare Eq. (3) of the main text. The self-energy $\Sigma = \Sigma_l + \Sigma_r$ arises from tunnel-coupling to left lead $\Sigma_l = t_l G_l t_l^\dagger$ and right lead $\Sigma_r = t_r G_r t_r^\dagger$ respectively, where $G_{l,\gamma}^{-1} = i\partial_t - \epsilon_\gamma$ and $G_{r,\gamma}^{-1} = i\partial_t - \epsilon_{\tilde\gamma}$. However, before integrating out the leads, we should introduce the counting field.

### B. Introduction of counting fields

Formulas for the charge current and its noise can be conveniently derived via the introduction of a counting field. Following Ref. [1], we introduce a counting field $\lambda(t)$ for the charge transported into the left lead $Q_l$ by adding $\mathcal{S}_c = -\oint_K dt\,\lambda(t)\sum_{\gamma\sigma}\bar\Psi_{\gamma\sigma}\Psi_{\gamma\sigma}$ to the action, that is, $\mathcal{S} \to \mathcal{S} + \mathcal{S}_c$. This newly added term is eliminated by a gauge transformation for the fermionic fields of the left lead: $\Psi_{\gamma\sigma} \to e^{-i\lambda(t)}\Psi_{\gamma\sigma}$ and $\bar\Psi_{\gamma\sigma} \to \bar\Psi_{\gamma\sigma} e^{i\lambda(t)}$. While this gauge transformation eliminates $\mathcal{S}_c$, it modifies the tunneling-matrix as

$$t_l \to t_l e^{-i\lambda(t)} \qquad \text{and} \qquad t_l^\dagger \to t_l^\dagger e^{+i\lambda(t)}\ . \tag{S1}$$

When the leads are integrated out, the counting field is transferred to the self-energy of the left lead $\Sigma_l(t-t') \to e^{-i\lambda(t)}\Sigma_l(t-t')e^{+i\lambda(t')}$. For compact notation, we write $\Sigma_l(\lambda) = e^{-i\lambda(t)}\Sigma_l(t-t')e^{+i\lambda(t')}$. We assume the counting field to have only a quantum component $\lambda^\pm(t) = \pm\lambda_q(t)/2$. For simplicity, we assume the counting field to be constant $\lambda_q(t) = \lambda$, which is possible as we are interested in the zero-frequency current and noise [2].

Now, we integrate out the leads as indicated above. Afterwards, we also integrate out the fermionic fields of the small ferromagnet to obtain the Keldysh partition function

$$\mathcal{Z}(\lambda) = e^{i\mathcal{S}(\lambda)} \tag{S2}$$

with the action

$$\mathcal{S}(\lambda) = -i \operatorname{tr} \ln \big[ \underbrace{i\partial_t - h_0 + \boldsymbol{M\sigma}/2 - \Sigma(\lambda)}_{G_\lambda^{-1}} \big] , \tag{S3}$$

where the counting field is contained in the self-energy $\Sigma = \Sigma_l(\lambda) + \Sigma_r$. The counting field was introduced in such a way that the charge transported through the left junction is determined as

$$\langle Q_l \rangle = i\partial_\lambda \; \mathcal{Z}(\lambda)|_{\lambda=0} \qquad \text{and} \qquad \langle Q_l^2 \rangle = (i\partial_\lambda)^2 \; \mathcal{Z}(\lambda)|_{\lambda=0} \tag{S4}$$

and analog for higher moments. Would we directly take the derivative with respect to the counting field $\lambda$, we would obtain for the charge $\langle Q_l \rangle = -i \operatorname{tr}[G_0 \Sigma_l']$ with $\Sigma_l' = \partial_\lambda \Sigma_l(\lambda)|_{\lambda=0}$ and for its noise $\langle\langle Q_l^2 \rangle\rangle = \operatorname{tr}[G_0 \Sigma_l''] + \operatorname{tr}[G_0' \Sigma_l']$ with $\Sigma_l'' = \partial_\lambda^2 \Sigma_l(\lambda)|_{\lambda=0}$ and $G_0' = \partial_\lambda G_\lambda\big|_{\lambda=0} = G_0 \Sigma_l' G_0$. The problem is, however, that we do not know the laboratory-frame Green's function $G_0$. This is the reason for making the transformation to the rotating frame, where we can determine the Green's function $\tilde{G}_0$.

### C. Derivation of results for charge current and its noise.

The counting field is chosen to have a quantum component only and to be constant in time, that is $\lambda^\pm(t) = \pm \frac{\lambda_q}{2}$. To keep the notation simple, we drop the quantum-index $\lambda_q \to \lambda$ in the following. After changing to the rotating frame with $R^\dagger \boldsymbol{M\sigma} R = M\sigma_z$, the generating function is given by $\mathcal{Z}(\lambda) = e^{i\mathcal{S}(\lambda)}$ with the new action

$$\mathcal{S}(\lambda) = -i \; \operatorname{tr} \ln \big[ \underbrace{i\partial_t - \epsilon_\alpha + M\sigma_z/2 - R^\dagger \Sigma(\lambda) R}_{\tilde{G}_\lambda^{-1}} \big] . \tag{S5}$$

Due to the time-dependence of the rotations, a new term $iR^\dagger \dot{R}$ arises. However, as in the main text, the spin-diagonal part of $iR^\dagger \dot{R}$ is eliminated by the choice of gauge $\dot{\chi} = \dot{\phi}(1-\cos\theta)$ and the spin-off-diagonal part of $iR^\dagger \dot{R}$ is disregarded in an adiabatic approximation.

Using equation (S4), a straightforward differentiation with respect to the counting field leads to the first moment $\langle Q_l \rangle = -i \operatorname{tr}\big[\tilde{G}_0 \tilde{\Sigma}_l'\big]$, where $\tilde{\Sigma}_l' = \partial_\lambda R^\dagger \Sigma_l(\lambda) R\big|_{\lambda=0}$. From the main text, we know the spin-diagonal parts of the rotated self-energy $\tilde{\Sigma}_{\sigma\sigma}^{R/A}(\omega) = \mp i\Gamma_l$ and $\tilde{\Sigma}_{\sigma\sigma}^K(\omega) = -2i\Gamma_l \tilde{F}_l^\sigma(\omega)$ and also the Green's function $\tilde{G}_0^{R/A}(\omega) \approx 1/(\omega - \xi_{a\sigma} \pm i\Gamma_\Sigma)$ and $\tilde{G}_0^K(\omega) \approx -2i\Gamma_\Sigma \tilde{F}_s^\sigma(\omega)/[(\omega - \xi_{a\sigma})^2 + \Gamma_\Sigma^2]$. The spin-off-diagonal parts can be disregarded due the large magnetization $M$. This leads to

$$\langle Q_l \rangle = \int dt \int d\omega \sum_\sigma \rho_\sigma \Gamma_l \big[\tilde{F}_l^\sigma(\omega) - \tilde{F}_s^\sigma(\omega)\big] , \tag{S6}$$

where we assumed the density of states $\rho_\sigma(\omega) = \sum_a \frac{\Gamma_\Sigma}{(\omega - \xi_{a\sigma})^2 + \Gamma_\Sigma^2}$ to be approximately constant on all scales smaller than $M$ around the leads' electrochemical potentials $\mu$, i.e., $\rho_\sigma(\mu + \omega) = \rho_\sigma$. With $\langle Q_l \rangle = \int dt I_l$, we can immediately read of the result for the current $I_l = \sum_\sigma \rho_\sigma \Gamma_l \int d\omega \big[\tilde{F}_l^\sigma(\omega) - \tilde{F}_s^\sigma(\omega)\big]$ which is the Landauer formula with rotating-frame distribution functions. The results for the right junction can be obtained analogously by introducing a counting field for the right lead. The formal result for $I_r$ is analog to $I_l$ but with the replacements $\Gamma_l \to \Gamma_r$ and $\tilde{F}_l^\sigma(\omega) \to \tilde{F}_r^\sigma(\omega)$. However, charge conservation demands $I_l + I_r = 0$ at zero frequency. Due to the absence of bias, we explicitly find $I_l = 0$ and $I_r = 0$ which could have been expected from symmetry.

For the second moment follows $\langle Q_l^2 \rangle = N_l^0 + N_l^1 + N_l^2$ with $N_l^0 = -\big(\operatorname{tr}\big[\tilde{G}_0 \tilde{\Sigma}_l'\big]\big)^2 = \langle Q_l \rangle^2$, such that we obtain the cumulant

$$\langle\langle Q_l^2 \rangle\rangle = \langle Q_l^2 \rangle - \langle Q_l \rangle^2 = N_l^1 + N_l^2 . \tag{S7}$$

Formally, $N_l^1 = \operatorname{tr}\big[\tilde{G}_0 \tilde{\Sigma}_l''\big]$ and $N_l^2 = \operatorname{tr}\big[\tilde{G}_0 \tilde{\Sigma}_l' \tilde{G}_0 \tilde{\Sigma}_l'\big]$, where $\tilde{\Sigma}_l'' = \partial_\lambda^2 R^\dagger \Sigma_l(\lambda) R\big|_{\lambda=0}$ and in $N_l^2$ we used $\partial_\lambda \tilde{G}_\lambda\big|_{\lambda=0} = \tilde{G}_0 \tilde{\Sigma}_l' \tilde{G}_0$. The term $N_l^2$ arises from the dependence of the Green's function on the counting field. Thus, this term can be interpreted as a reaction of the distribution function of the small ferromagnet to tunneling of electrons. One might wounder, if it is important to take those reactions of the distribution function into account. The answer is a clear yes. Would we approximate $\langle\langle Q_l^2 \rangle\rangle \approx N_l^1$ and proceeding analogously for the right contact $\langle\langle Q_r^2 \rangle\rangle \approx N_r^1$, we would

find $\langle\langle Q_l^2\rangle\rangle \neq \langle\langle Q_r^2\rangle\rangle$ which violates charge conservation, as we consider zero-frequency. More explicitly, we obtain

$$N_l^1 = \int dt \int d\omega \sum_\sigma \rho_\sigma(\omega)\, \Gamma_l \big[1 - \tilde{F}_s^\sigma(\omega)\tilde{F}_l^\sigma(\omega)\big] \tag{S8}$$

$$N_l^2 = \int dt \int d\omega \sum_\sigma \frac{\Gamma_l^2}{\Gamma_\Sigma} \Big[2\bar{\rho}_\sigma(\omega)\tilde{F}_s^\sigma(\omega)\tilde{F}_l^\sigma(\omega) - \bar{\rho}_\sigma(\omega)\tilde{F}_s^\sigma(\omega)\tilde{F}_s^\sigma(\omega) - \rho_\sigma(\omega) - 2[\tilde{F}_l^\sigma(\omega)]^2(\bar{\rho}_\sigma(\omega) - \rho_\sigma(\omega))\Big]\,, \tag{S9}$$

with two differently broadened densities of states $\rho_\sigma(\omega) = \sum_a \frac{1}{\pi}\frac{\Gamma_\Sigma}{(\omega-\xi_{a\sigma})^2+\Gamma_\Sigma^2}$ and $\bar{\rho}_\sigma(\omega) = \sum_a \frac{1}{\pi}\frac{2\Gamma_\Sigma^3}{((\omega-\xi_{a\sigma})^2+\Gamma_\Sigma^2)^2}$. We assume this difference in broadenings to be insignificant, that is, we approximate $\bar{\rho}_\sigma(\omega) \approx \rho_\sigma(\omega)$. As for the current, we assume the density of states to be approximately constant $\rho_\sigma(\omega) \approx \rho_\sigma$. It follows,

$$\langle\langle Q_l^2\rangle\rangle = \int dt \int d\omega \sum_\sigma \frac{\rho_\sigma \Gamma_r \Gamma_l}{\Gamma_\Sigma} \left\{ \big[1 - \tilde{F}_s^\sigma(\omega)\tilde{F}_l^\sigma(\omega)\big] + \frac{\Gamma_l}{\Gamma_r}\tilde{F}_s^\sigma(\omega)\big[\tilde{F}_l^\sigma(\omega) - \tilde{F}_s^\sigma(\omega)\big] \right\}\,. \tag{S10}$$

From this result, we can read off the noise of charge current $S_l$, which is defined by $\langle\langle Q^2\rangle\rangle = \int dt\, S_l/2$. And indeed, with $N_l^1$ and $N_l^2$ both taken into account, charge conservation is satisfied $\langle\langle Q_l^2\rangle\rangle = N_l^1 + N_l^2 = N_r^1 + N_r^2 = \langle\langle Q_r^2\rangle\rangle$, when we proceed analogously for the right contact.

### S.II. LABORATORY-FRAME DISTRIBUTION FUNCTION WOULD BE HARD TO DETERMINE

The rotating-frame distribution functions of the leads $\tilde{F}_{l/r}(t,t')$ have been found by the rotation of the laboratory-frame distribution functions $F_{l/r}(t-t')$, that is, $\tilde{F}_{l/r}(t,t') = R^\dagger(t)F_{l/r}(t-t')R(t')$. We could also perform the inverse rotation to obtain $F_{l/r}(t-t') = R(t)\tilde{F}_{l/r}(t,t')R^\dagger(t')$, which determines the leads' laboratory-frame distribution functions in terms of their rotating-frame distribution functions. Analogously, the laboratory-frame distribution function of the small ferromagnet is determined by $F_s(t,t') = R(t)\tilde{F}_s(t,t')R^\dagger(t')$, where $\tilde{F}_s(t,t')$ is the rotating-frame distribution function. With spin-indices written explicitly follows

$$F_s^{\sigma\sigma'''}(t,t') = \sum_{\sigma'\sigma''} R_{\sigma\sigma'}(t)\tilde{F}_s^{\sigma'\sigma''}(t,t')[R^\dagger(t')]_{\sigma''\sigma'''}\,. \tag{S11}$$

Using the $\theta = const.$ and $\dot{\phi} = const.$, the spin-diagonal contributions of the laboratory-frame distribution function become

$$F_s^{\uparrow\uparrow}(\bar{t},\omega) = \cos^2\frac{\theta}{2}\tilde{F}_s^{\uparrow\uparrow}(\omega-\omega_-) + \sin^2\frac{\theta}{2}\tilde{F}_s^{\downarrow\downarrow}(\omega-\omega_+) - \frac{\sin\theta}{2}\Big[\tilde{F}_s^{\downarrow\uparrow}(\omega-\dot{\phi}/2)e^{-i\dot{\phi}\cos\theta\,\bar{t}} + \tilde{F}_s^{\uparrow\downarrow}(\omega-\dot{\phi}/2)e^{i\dot{\phi}\cos\theta\,\bar{t}}\Big]\,, \tag{S12}$$

$$F_s^{\downarrow\downarrow}(\bar{t},\omega) = \cos^2\frac{\theta}{2}\tilde{F}_s^{\downarrow\downarrow}(\omega+\omega_-) + \sin^2\frac{\theta}{2}\tilde{F}_s^{\uparrow\uparrow}(\omega+\omega_+) + \frac{\sin\theta}{2}\Big[\tilde{F}_s^{\downarrow\uparrow}(\omega+\dot{\phi}/2)e^{-i\dot{\phi}\cos\theta\,\bar{t}} + \tilde{F}_s^{\uparrow\downarrow}(\omega+\dot{\phi}/2)e^{i\dot{\phi}\cos\theta\,\bar{t}}\Big]\,, \tag{S13}$$

where $\bar{t} = (t+t')/2$. However, despite this formal result, it is hard to make further progress. While we know the spin-diagonal parts of the rotating-frame distribution function, $\tilde{F}_s^{\uparrow\uparrow}(\omega)$ and $\tilde{F}_s^{\downarrow\downarrow}(\omega)$, we do not know the spin-off-diagonal parts $\tilde{F}_s^{\downarrow\uparrow}(\omega)$ and $\tilde{F}_s^{\uparrow\downarrow}(\omega)$. The spin-diagonal parts are hard to determine, because of the time-dependence of the spin-off-diagonal contributions of the rotating-frame self-energy and the retarded and advanced part of the rotating-frame Green's functions. So, we cannot determine the spin-diagonal parts of the laboratory-frame distribution functions.

In strong contrast to the distribution function $F_s$, it is straightforward to determine the laboratory-frame Keldysh Green's function $G^K$ from its rotating-frame version $\tilde{G}^K$. The inverse rotation $G^K(t,t') = R(t)\tilde{G}^K(t,t')R^\dagger(t')$ is analog to to the distribution function $F_s$. With spin-indices written out explicitly we obtain,

$$G_{\sigma\sigma'''}^K(t,t') = \sum_{\sigma'\sigma''} R_{\sigma\sigma'}(t)\tilde{G}_{\sigma'\sigma''}^K(t,t')[R^\dagger(t')]_{\sigma''\sigma'''}\,. \tag{S14}$$

However, in strong contrast to the distribution function, we know that the spin-off-diagonal contributions to the rotating-frame Keldysh Green's function $\tilde{G}_{\sigma\bar\sigma}^K = \tilde{G}_\sigma^R \tilde{\Sigma}_{\sigma\bar\sigma}^K \tilde{G}_{\bar\sigma}^A$ are suppressed by the large value of the magnetization $M$. Therefore, it is sufficient to take into account only the spin-diagonal elements. A straightforward calculation yields,

$$G_{\sigma\sigma}^K(\omega) = \cos^2\frac{\theta}{2}\tilde{G}_{\sigma\sigma}^K(\omega-\sigma\omega_-) + \sin^2\frac{\theta}{2}\tilde{G}_{\sigma\sigma}^K(\omega-\sigma\omega_+)\,, \tag{S15}$$

$$G_{\sigma\bar\sigma}^K(\bar{t},\omega) = \frac{\sin\theta}{2}\big[\tilde{G}_{\uparrow\uparrow}^K(\omega+\cos\theta\,\dot{\phi}/2) - \tilde{G}_{\downarrow\downarrow}^K(\omega-\cos\theta\,\dot{\phi}/2)\big]e^{-i\sigma\dot{\phi}\bar{t}}\,, \tag{S16}$$

with the spin-diagonal part of the rotating-frame Keldysh Green's function $\tilde{G}_0^K(\omega) \approx -2i\Gamma_\Sigma \tilde{F}_s^\sigma(\omega)/[(\omega-\xi_{a\sigma})^2+\Gamma_\Sigma^2]$.

---



\end{document}